\pdfoutput=1
\documentclass[aps,prl,superscriptaddress,twocolumn,floatfix]{revtex4-1}

\usepackage{appendix}
\usepackage{graphicx,graphics,grffile}% Include figure files
\usepackage{dcolumn}% Align table columns on decimal point
\usepackage{bm}% bold math
\usepackage{braket}
\usepackage{nicefrac}
\usepackage{feynmp}
\usepackage{siunitx}
\usepackage[utf8]{inputenc}
\usepackage[abs]{overpic}
\usepackage{dcolumn}
\usepackage{amsmath,amssymb,amsfonts}
\usepackage{wasysym}
\usepackage{latexsym,verbatim}
\usepackage{color}
\usepackage[framemethod=default]{mdframed}
\usepackage[breaklinks=true,colorlinks,citecolor=blue,linkcolor=blue,urlcolor=blue]{hyperref}
\usepackage[makeroom]{cancel}
\usepackage{fancybox}

\usepackage{natbib}
\usepackage{array,booktabs}

\usepackage[abs]{overpic}

\usepackage{mathrsfs}

\newcommand{\be}{\begin{eqnarray}}
\newcommand{\ee}{\end{eqnarray}}
\newcommand{\nn}{\nonumber\\}

\begin{document}

\title{Majorana corner states in square and Kagome quantum spin liquids}
\author{Haoran Wang}
%\email{}
\affiliation{\noindent Department of Physics and Astronomy, University of Manchester, Manchester M13 9PL, UK}
\author{Alessandro Principi}
\email{alessandro.principi@manchester.ac.uk}
\affiliation{\noindent Department of Physics and Astronomy, University of Manchester, Manchester M13 9PL, UK}

\begin{abstract}
Quantum spin liquids hosting Majorana excitations have recently experienced renewed interest for potential applications to topological quantum computation. Performing logical operations with reduced poisoning requires to localize such quasiparticles at specific point of the system, with energies that are well defined and inside the bulk energy gap. These are two defining features of second order topological insulators (SOTIs). Here, we show two spin models that support quantum spin liquid phases characterised by Majorana excitations and that behave as SOTIs, one of which is analytically solvable thanks to a theorem by Lieb. We show that, depending on the values of spin couplings, it is possible to localize either fermions or Majorana particles at their corners.
\end{abstract}

\maketitle

{\it Introduction}---Quantum spin liquids (QSLs) are intriguing states of matter in which spins never freeze due to their high degree of entanglement~\cite{Anderson_mrb_1973,Zhou_rmp_2017,Savary_rpp_2016,Wen_book}. Recently, they have experienced rekindled interest because of two main factors. On the one hand, the development of exactly-solvable spin-lattice models, for example Kitaev's~\cite{Kitaev_AnnPhys_2006}, which host QSL phases whose excitations are neither fermions nor bosons but Majorana particles. On the other hand, the discovery of new materials compatible with such models~\cite{Banerjee_science_2017}. 

Majorana particles are one of the cornerstones of research in topological quantum computation, since they can be used to design quantum gates that are resilient to external noise~\cite{Kitaev_AnnPhys_2003,Sarma_npj_2015,Hoffman_prb_2016,Lian_pnas_2018}. 
In order to use them to perform logical operations, it is however crucial to find ways to localize them at specific points of the system, with well-defined energies inside an energy gap. The former property is missing in the original Kitaev QSL model. There, 1D Majorana channels can be formed at the edges, in the presence of a magnetic field~\cite{Kitaev_AnnPhys_2006}. (In principle, it is also possible to localize Majorana particles at vortices of the $\mathbb{Z}_2$ gauge field~\cite{Kitaev_AnnPhys_2006,Knolle_thesis}. However, generating and controlling such nanoscale objects in experiments would be extremely challenging.) In this paper we show two spin models, one of which analytically solvable, which support topologically-protected Majorana {\it corner} states with energies in the middle of the bulk band gap.

Mid-gap corner states emerge in  two-dimensional (2D) ``second-order'' topological insulators (SOTIs)~\cite{Benalcazar_Science_2017,Benalcazar_prb_2017,Schindler_SciAdv_2018,Langbehn_prl_2017,Geier_prb_2018,Ezawa_prl_2018,Song_prl_2017,Ezawa_prb_2018,Ezawa_prb_2018b}. $d$-dimensional SOTIs are insulating both in the bulk and at the surfaces, but exhibit gapless $d-2$-states protected by a variety of crystalline symmetries~\cite{Benalcazar_prb_2017} (among others, mirror reflection, twofold rotation, or inversion symmetry). SOTIs have been realized in various experiments, most notably in artificial settings such as mechanical~\cite{Serra-Garcia_nature_2018}, electrical~\cite{Imhof_natphys_2018}, microwave~\cite{Peterson_nature_2018}, and photonic~\cite{Hassan_nature_2019} devices, but they have also been shown to occur in Nature~\cite{Schindler_natphys_2018}. A similar phenomenology has been predicted to occur in non-Hermitian systems~\cite{Luo_prl_2019}, topological superconductors~\cite{Liu_prb_2018,Zhu_prb_2018,Laubscher_prr_2019,Wang_prl_2018,Kheirkhah_prl_2020,Yan_prl_2019} and QSLs~\cite{Dwivedi_prb_2018}. In the latter two, corner states can be Majorana particles. However, to the best of our knowledge, no analytically solvable QSL has been shown to exhibit Majorana corner states~\footnote{The model of Ref.~\cite{Dwivedi_prb_2018} can be solved exactly by mapping spins into Majorana particles which are free to propagate on top a quenched $\mathbb{Z}_2$ gauge potential. However, the ground-state flux configuration is not analytically known and must be found with numerical techniques.}.

Here we study two frustrated spin-$3/2$ systems which exhibit QSL phases and corner states at low temperature. The models we study are similar to that of Ref.~\cite{Dwivedi_prb_2018} but are defined on Kagome~\cite{Chua_prb_2011} and square~\cite{Yao_prl_2009} lattices, rather than on the Shastry-Sutherland one. Following Kitaev's construction~\cite{Kitaev_AnnPhys_2006}, we fractionalize the spin degrees of freedom into two itinerant Majorana particles and a quesched $\mathbb{Z}_2$ gauge potential. Notably, the ground state of the square lattice we study is analytically known~\cite{Lieb_prl_1994}. We show that both models can support stable fermionic or Majorana corner states, depending on the values of the spin couplings. The analytical solvability of the square-lattice model lends credibility to the possibility of finding Majorana corner states in QSLs.

{\it The model}---We consider two lattices, square and Kagome, at whose sites are located spin-$3/2$ variables. The Hamiltonian of such systems can be written in terms of $4\times 4$ matrices which operate on the four spin polarizations at each site. The basis of $4\times 4$ matrices can be chosen to be composed by: the identity, five Gamma matrices ${\hat \Gamma}^a$ ($a=1,\ldots,5$) which can be represented as symmetric bilinear combinations of the spin-$3/2$ operators ${\hat S}^\alpha$ ($\alpha = x,y,z$) as~\cite{Chua_prb_2011,Yao_prl_2009}
\be
{\hat \Gamma}^1 = \frac{\{{\hat S}^y, {\hat S}^z\}}{\sqrt{3}},~  {\hat \Gamma}^2 = \frac{\{{\hat S}^z, {\hat S}^x\}}{\sqrt{3}},~  {\hat \Gamma}^3 = \frac{\{{\hat S}^x, {\hat S}^y\}}{\sqrt{3}}, 
\nonumber
\ee
\be
{\hat \Gamma}^4 = \frac{1}{\sqrt{3}} \big[({\hat S}^x)^2 - ({\hat S}^y)^2\big],~ {\hat \Gamma}^5 = ({\hat S}^z)^2 - \frac{5}{4},
\ee
and the ten bilinears ${\hat \Gamma}^{ab} = [{\hat \Gamma}^a, {\hat \Gamma}^b]/(2i)$. Here, $[\cdot,\cdot]$ and $\{\cdot,\cdot\}$ are the matrix commutator and anticommutator, respectively. In passing we note that the chosen Gamma matrices satisfy the Clifford algebra $\{{\hat \Gamma}^a,{\hat \Gamma}^b\} = 2 \delta^{ab}$.

\begin{figure}[t]
\begin{center}
\begin{tabular}{cc}
\begin{overpic}[width=0.39\columnwidth]{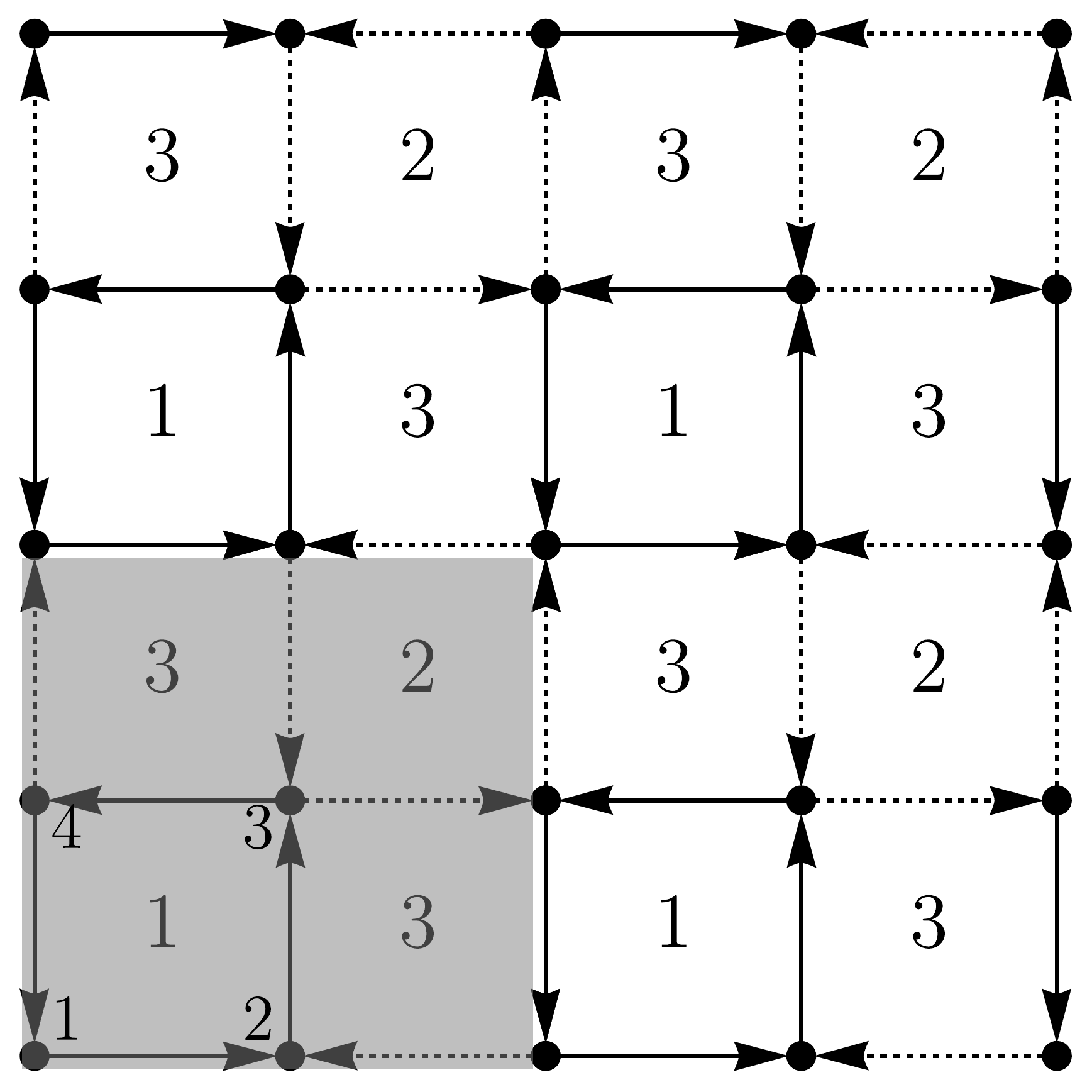}
\put(0,-10){(a)}
\end{overpic}
&
\begin{overpic}[width=0.59\columnwidth]{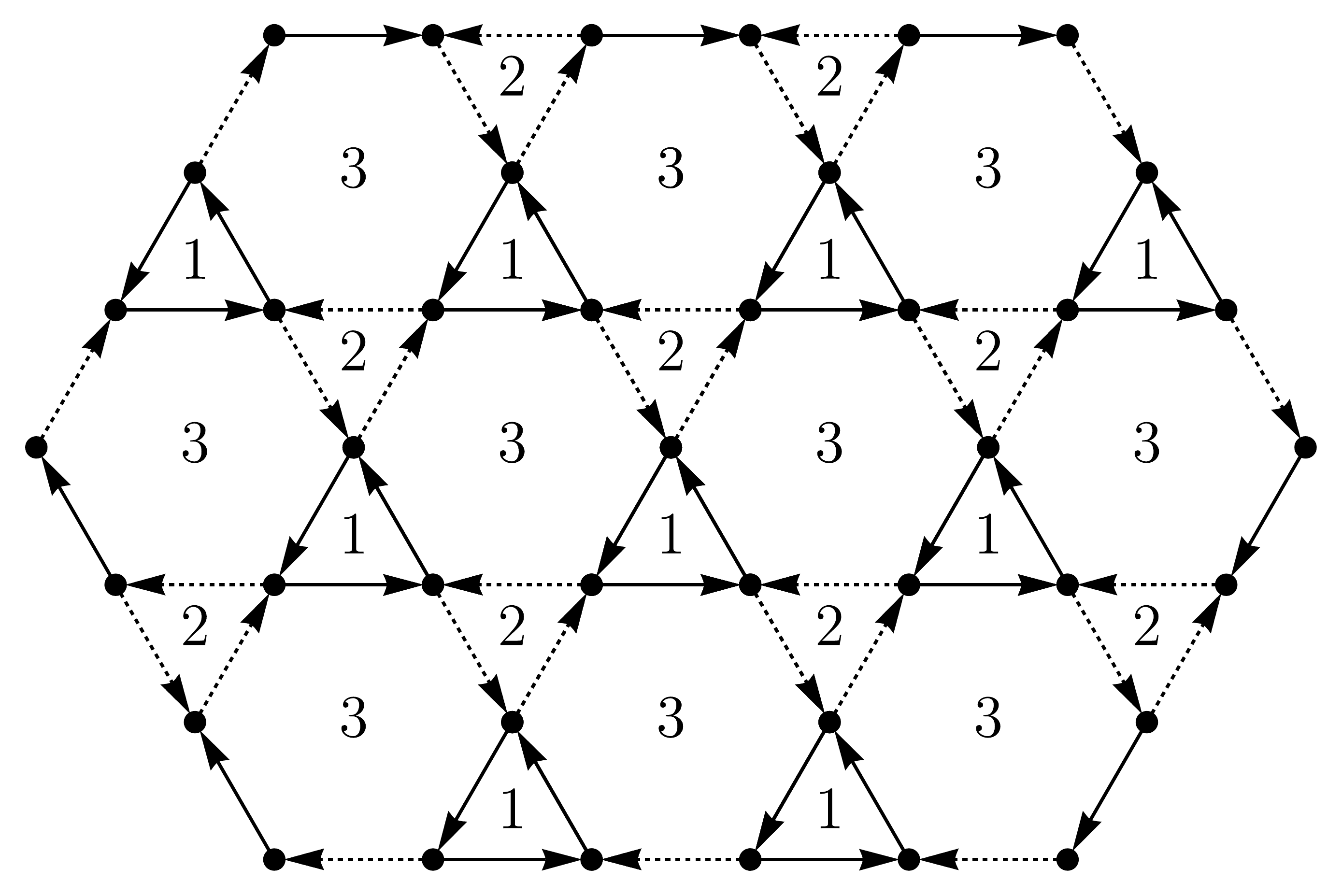}
\put(0,-10){(b)}
\end{overpic}
\end{tabular}
\vspace{0.5cm}\\
\begin{tabular}{cccc}
\begin{overpic}[width=0.24\columnwidth]{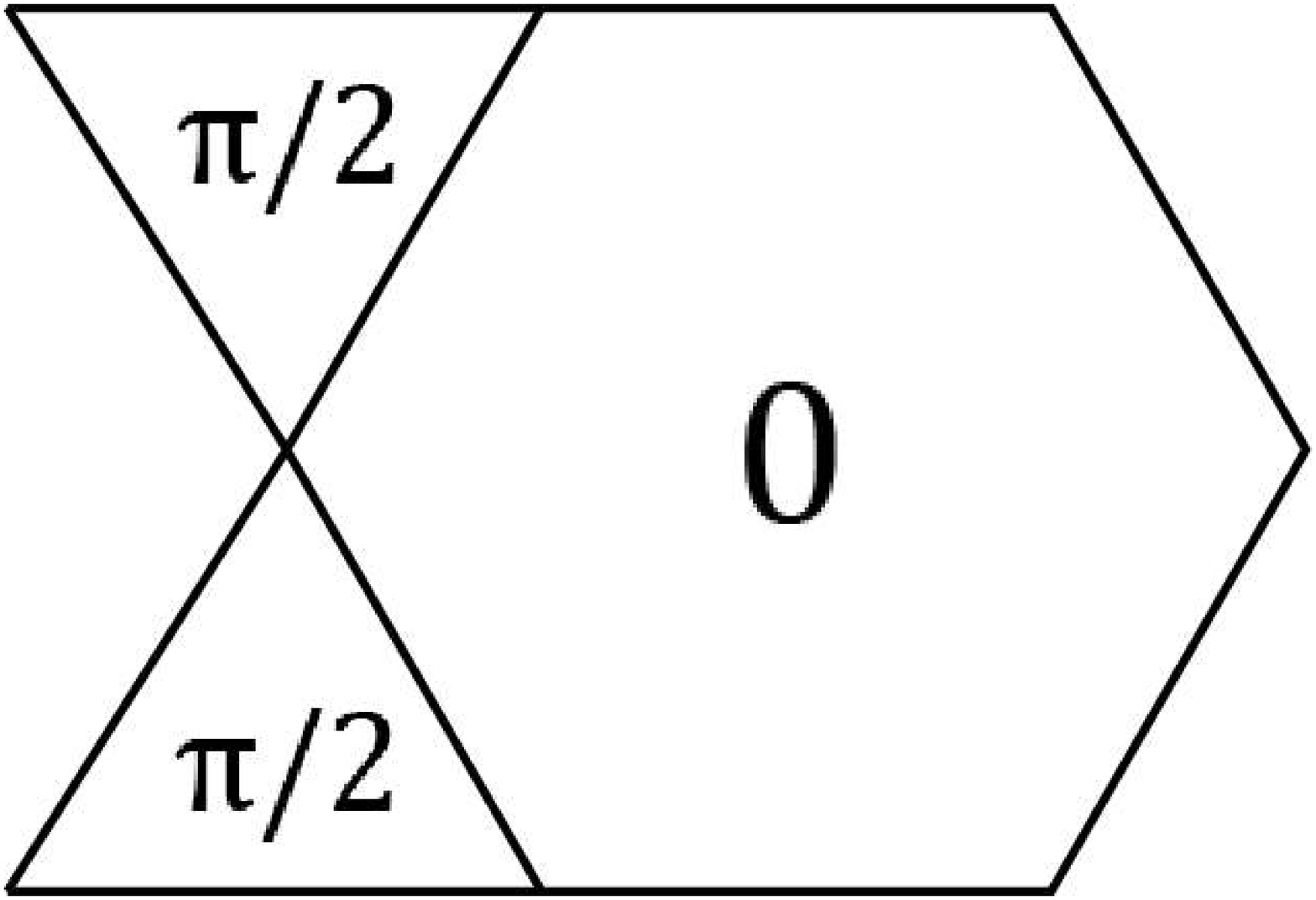}
\put(0,-10){(c)}
\end{overpic}
&
\begin{overpic}[width=0.24\columnwidth]{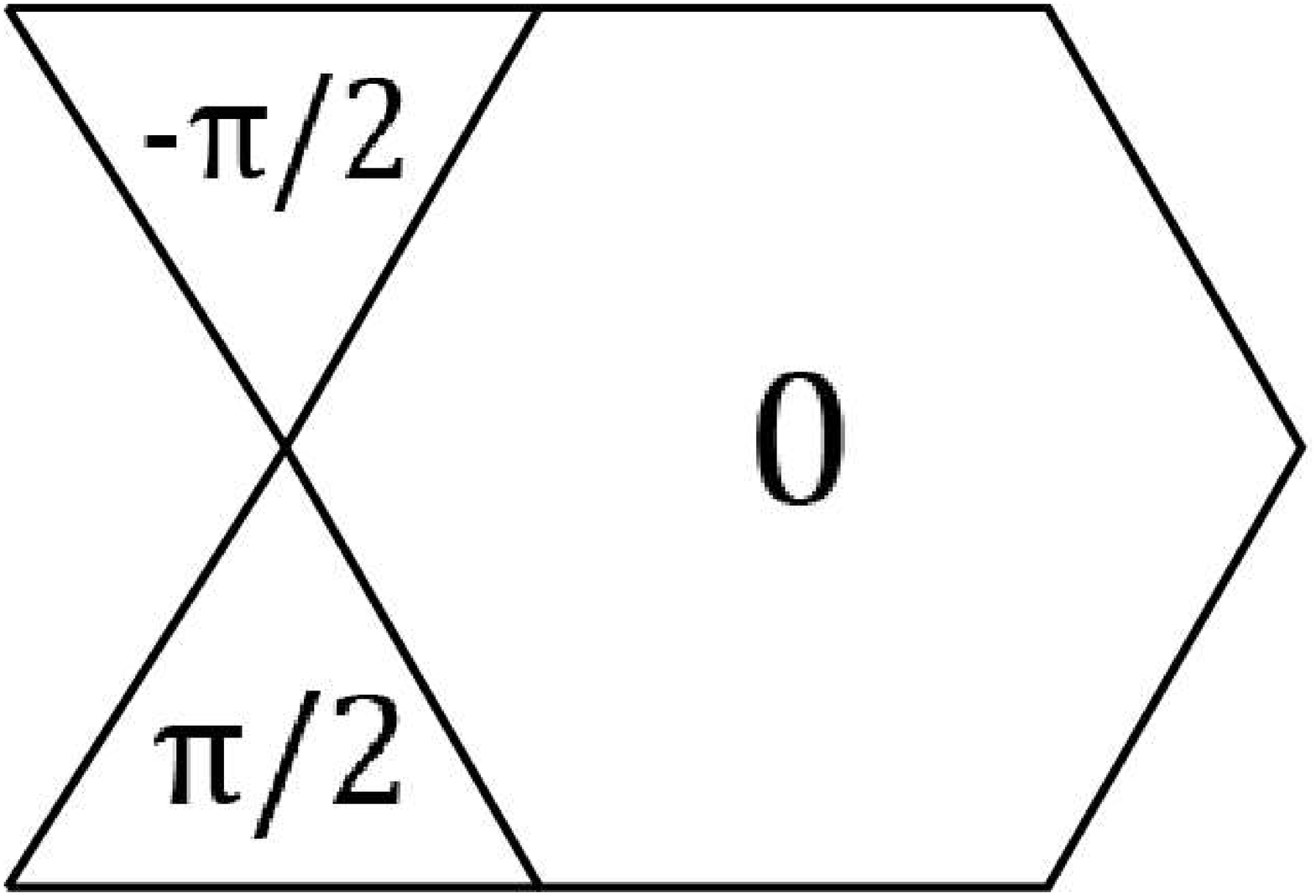}
\put(0,-10){(d)}
\end{overpic}
&
\begin{overpic}[width=0.24\columnwidth]{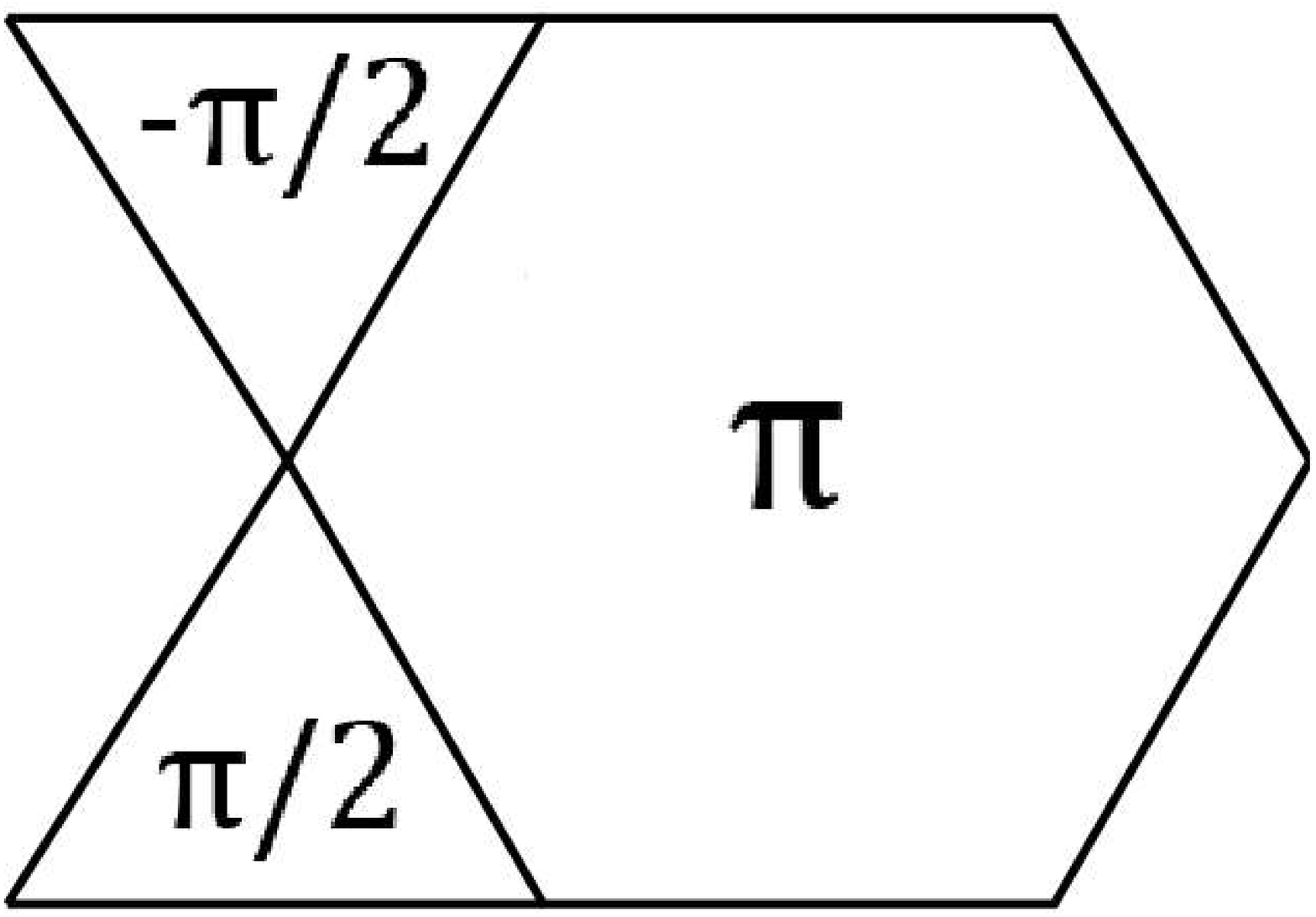}
\put(0,-10){(e)}
\end{overpic}
&
\begin{overpic}[width=0.24\columnwidth]{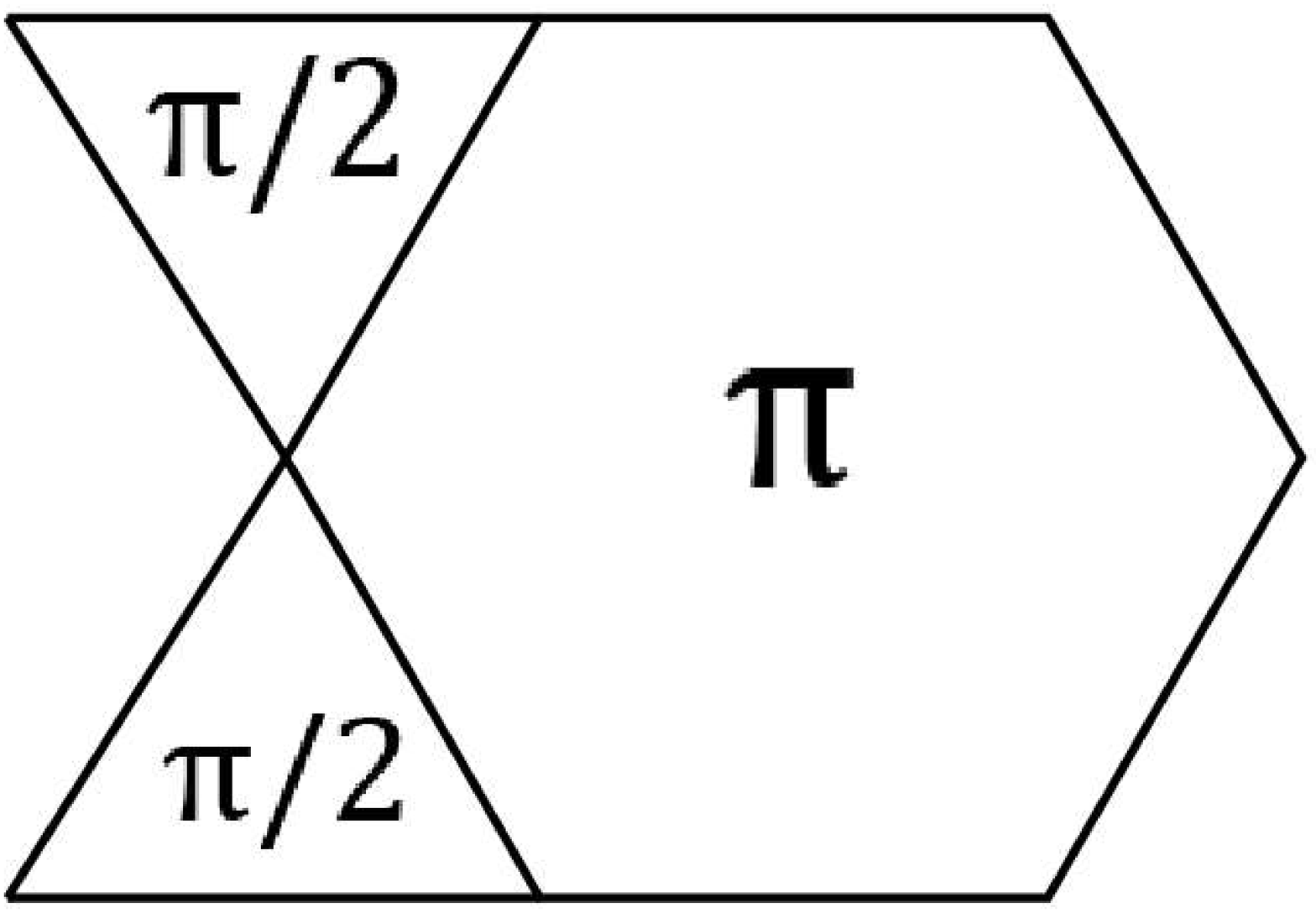}
\put(0,-10){(f)}
\end{overpic}
\end{tabular}
\end{center}
\caption{A pictorial representation of the two models studied in this paper. 
Panel (a) The square model.
Panel (b) The Kagome model.
In both cases we identify three types of plaquettes and two types of bonds. Solid (dashed) arrows correspond to couplings of the type ${\hat \Gamma}_i^1{\hat \Gamma}_j^2$ and ${\hat \Gamma}_i^{15}{\hat \Gamma}_j^{25}$ (${\hat \Gamma}_i^3{\hat \Gamma}_j^4$ and ${\hat \Gamma}_i^{35}{\hat \Gamma}_j^{45}$). The direction of the arrow is from site $i$ to $j$. The shaded region represents the unit cell used in square-lattice calculations (atoms in it are labeled with $1,2,3$ and $4$).
Panels (c)-(f) The four distinct flux patterns in the Kagome unit cell. Panels (c) and (e) require doubling of the unit cell. 
}
\label{fig:one}
\end{figure}

We now define the Gamma-matrix model~\cite{Chua_prb_2011,Yao_prl_2009}
\be \label{eq:Gamma_model_Hamiltonian}
{\hat {\cal H}} &=& 
J_1 \sum_{\langle i,j \rangle\in {\cal P}_1} {\hat \Gamma}_i^1{\hat \Gamma}_j^2+ J_2 \sum_{\langle i,j \rangle\in {\cal P}_2} {\hat \Gamma}_i^3{\hat \Gamma}_j^4
\nn
&+&
J_1' \sum_{\langle i,j \rangle\in {\cal P}_1} {\hat \Gamma}_i^{15}{\hat \Gamma}_j^{25} + J_2' \sum_{\langle i,j \rangle \in {\cal P}_2} {\hat \Gamma}_i^{35}{\hat \Gamma}_j^{45}
+
J_5\sum_i{\hat \Gamma}_i^5,
\nn
\ee
where $i$ and $j$ label the lattice sites, while ${\cal P}_1$ and ${\cal P}_2$ are the collections of plaquettes of types 1 and 2, respectively (see Fig.~\ref{fig:one} for the definition). In Eq.~(\ref{eq:Gamma_model_Hamiltonian}), the coupling between spins depends on the type of plaquette to which the link $\langle i, j\rangle$ belongs (links are taken in the counterclockwise direction in plaquettes 1 and 2). For the Kagome lattice, plaquettes 1 and 2 coincide with upward and downwards triangles, respectively. Conversely, the plaquettes of the square lattice are taken to alternate between type 1 and 2 along every other diagonal, while all others are of type 3. In the Kagome lattice, hexagons are type-3 plaquettes. 
Type-3 plaquettes do not appear explicitly in the Hamiltonian~(\ref{eq:Gamma_model_Hamiltonian}) to avoid double counting the bonds

It is possible to define a set of flux operators ${\hat W}_p$, one per plaquette $p$, which commute with the Hamiltonian and among themselves~\cite{Chua_prb_2011,Yao_prl_2009}. These are ${\hat W}_p = \prod_{i\in p} i{\hat \Gamma}^{12}_i$, if $p$ is of type 1, ${\hat W}_p = \prod_{i\in p} i {\hat \Gamma}^{34}_i$, if $p$ is of type 2, and ${\hat W}_p = \prod_{\langle i, j\rangle \in p} i{\hat \Gamma}^{23}_i {\hat \Gamma}^{14}_j$, if $p$ is of type 3. All products run over links taken in the counterclockwise direction around a given plaquette. In the type-3 flux operator, the first link is shared with a plaquette of type 1.
The presence of such a large number of constants of motion is at the root of the solvability of the models.

We now introduce six Majorana operators~\cite{Chua_prb_2011,Yao_prl_2009} ${\hat \xi}_i^1,{\hat \xi}_i^2,{\hat \xi}_i^3,{\hat \xi}_i^4,{\hat c}_i,$ and ${\hat d}_i$ at each site $i$, which satisfy the anticommutation relations $\{{\hat \xi}_i^a, {\hat \xi}_j^b\} = 2\delta_{ij}\delta^{ab}$, $\{{\hat c}_i, {\hat c}_j\} = 2\delta_{ij}$, $\{{\hat d}_i, {\hat d}_j\} = 2\delta_{ij}$, and $\{{\hat c}_i, {\hat \xi}_j^a\} = \{{\hat d}_i, {\hat \xi}_j^a\} = 0$. The Gamma matrices are then expressed as ${\hat \Gamma}_i^a=i{\hat \xi}_i^a {\hat c}_i$, ${\hat \Gamma}_i^{a5}=i{\hat \xi}_i^a {\hat d}_i$ and ${\hat \Gamma}_i^5=i {\hat c}_i {\hat d}_i$,
where $a=1,2,3,4$. This representation enlarges the Hilbert space, introducing non-physical states that must be projected out at the end of the calculation~\cite{Chua_prb_2011,Yao_prl_2009,Kitaev_AnnPhys_2006}. To define the projection operator, we note that ${\hat D}_i = - {\hat \Gamma}_i^1 {\hat \Gamma}_i^2 {\hat \Gamma}_i^3 {\hat \Gamma}_i^4 {\hat \Gamma}_i^5 = 1$. In the Majorana representation, however, the eigenvalues of ${\hat D}_i = -i {\hat \xi}_i^1 {\hat \xi}_i^2 {\hat \xi}_i^3 {\hat \xi}_i^4 {\hat c}_i {\hat d}_i$ are $\pm 1$. For any physical state $|\Psi\rangle_{\rm phys}$, it must then be ${\hat D}_i |\Psi\rangle_{\rm phys}= |\Psi\rangle_{\rm phys}$, and therefore one can define the projection operator onto the physical Hilbert subspace as~\cite{Chua_prb_2011,Yao_prl_2009} ${\hat P} = \prod_i(1+{\hat D}_i)/2$.

In the Majorana representation, Eq.~(\ref{eq:Gamma_model_Hamiltonian}) becomes
\begin{equation} \label{eq:Hamiltonian_Majorana}
{\hat {\cal H}_{\rm M}}=i \sum_{\langle i, j\rangle \in {\cal P}_\alpha}[J_{\alpha} {\hat u}_{ij}^\alpha {\hat c}_i{\hat c}_j+J_{\alpha}' {\hat u}_{ij}^\alpha {\hat d}_i{\hat d}_j]+iJ_5\sum_i {\hat c}_i{\hat d}_i,
\end{equation}
where $\alpha = 1,2$, ${\hat u}_{ij}^1=-i{\hat \xi}_i^1{\hat \xi}_j^2$, and ${\hat u}_{ij}^2=-i{\hat \xi}_i^3{\hat \xi}_j^4$. It can be shown~\cite{Chua_prb_2011,Yao_prl_2009,Kitaev_AnnPhys_2006} that all ${\hat u}_{ij}^\alpha$ commute with the Hamiltonian~(\ref{eq:Hamiltonian_Majorana}). Consequently, the full Hilbert space can be divided into sectors, each obtained by replacing ${\hat u}_{ij}^\alpha$ with the eigenvalues $u_{ij}^\alpha = \pm 1$. In each sector, Eq.~(\ref{eq:Hamiltonian_Majorana}) describes free Majorana particles (${\hat c}$ and ${\hat d}$) propagating on top of a quenched $\mathbb{Z}_2$ gauge potential $u_{ij}^\alpha$. The eigenvalues of flux operators are $W_p = \prod_{\langle i, j \rangle \in p} i u_{ij}^\alpha \equiv e^{i \phi_p}$.
Here, $\phi_p$ is the flux through a given plaquette. For Kagome lattices, $\phi_p=\pm\pi/2$, if $p$ is of type 1 or 2, while $\phi_p=0$ or $\pi$ if $p$ is of type 3. Because of the antisymmetry of the Hamiltonian, a total of four distinct flux patterns exist for any unit cell [see Fig.~\ref{fig:one}(c)-(f)]. The total flux through a unit cell is either $0$ or $\pi$, and in the $\pi$ case doubling the unit cell is required. In the square lattice, the flux through any plaquette can only be either $\phi_p = 0$ or $\pi$. 

Not all sectors describe different physical systems. In fact, the variables commuting with the original spin Hamiltonian are the fluxes ${\hat W}_p$ and not the gauge potential ${\hat u}_{ij}^\alpha$.  By performing the gauge transformation ${\hat c}_i \to \Lambda_i {\hat c}_i$, ${\hat d}_i \to \Lambda_i {\hat d}_i$ and $u_{ij} \to \Lambda_i u_{ij}^\alpha \Lambda_j$, with $\Lambda_i = \pm 1$, both Hamiltonian and fluxes remain invariant. Hence, $2^N$ configurations of $u_{ij}^\alpha$ describe the same physical state~\cite{Chua_prb_2011,Yao_prl_2009,Kitaev_AnnPhys_2006}. One can choose to work with any configuration, depending on convenience: upon projection with ${\hat P}$, the physical state becomes a superposition of states of all equivalent Hilbert space sectors. 
We now set $J_5=0$ and analyze the two lattices separately.

{\it Square lattice}---We rewrite the Hamiltonian~(\ref{eq:Hamiltonian_Majorana}) as
\be \label{eq:Hamiltonian_Majorana_square}
{\hat {\cal H}_{\rm M}}&=& i \sum_{\ell, m}(-)^{\ell+m} \big[
u_{\ell,m}^x ({\tilde J}_{m} {\hat c}_{\ell, m} {\hat c}_{\ell, m+1} + {\tilde J}_{m}' {\hat d}_{\ell, m} {\hat d}_{\ell, m+1}) 
\nn
&-&
u_{\ell,m}^y ({\tilde J}_{\ell} {\hat c}_{\ell, m} {\hat c}_{\ell+1, m} + {\tilde J}_{\ell}' {\hat d}_{\ell, m} {\hat d}_{\ell+1, m}) 
\big],
\ee
where $\ell$ and $m$ denote the row and column in the lattice of Fig.~\ref{fig:one}(a), respectively, and $2{\tilde J}_{\ell} = (J_1 + J_2) + (-)^\ell (J_1 - J_2)$ (${\tilde J}_{\ell}'$ is analogously defined, with $J_1'$ and $J_2'$ in lieu of  $J_1$ and $J_2$). In Eq.~(\ref{eq:Hamiltonian_Majorana_square}), $u_{\ell,m}^x$ [$u_{\ell,m}^y$] is the value of the $\mathbb{Z}_2$ gauge field between sites $i = (\ell,m)$ and $j = (\ell +1, m)$ [$i = (\ell,m)$ and $j = (\ell, m +1)$] along the $x$ [$y$] direction. Thanks to Lieb's theorem~\cite{Lieb_prl_1994}, the ground state of the square lattice with equal hopping amplitudes is known to contain one flux quantum per plaquette ($\phi_p = \pi$)~\footnote{Strictly speaking, Lieb's theorem holds for periodic systems. Following Kitaev~\cite{Kitaev_AnnPhys_2006}, we will assume that the ground state of a large open lattice coincides with that of the associated periodic system.}. When the difference between hopping amplitudes is much smaller than the two-flux excitation energy, we expect the ground state configuration to remain unchanged.

We observe that $-{\hat {\cal H}_{\rm M}}$ has the same flux pattern as ${\hat {\cal H}_{\rm M}}$, since the latter is defined modulo $2\pi$. Therefore, $-{\hat {\cal H}_{\rm M}}={\hat {\cal G}}{\hat {\cal H}_{\rm M}}{\hat {\cal G}}^{-1}$, where ${\hat {\cal G}}$ is a gauge transformation that inverts the signs of all $u_{\ell,m}^{x(y)}$. This implies that the eigenvalues of ${\hat {\cal H}_{\rm M}}$ must be symmetric about zero, since if the state $|\psi\rangle$ has energy $E$, then the state ${\hat {\cal G}}|\psi\rangle$ has energy $-E$. 
We also note that, if ${\hat {\cal R}}$ is a $90^\circ$ rotation, there exist a gauge operation ${\hat {\cal G}}$ such that ${\hat {\cal R}}{\hat {\cal H}_{\rm M}}{\hat {\cal R}}^{-1} = {\hat {\cal G}}^{-1}{\hat {\cal H}_{\rm M}}{\hat {\cal G}}$. Therefore ${\hat {\cal G}}{\hat {\cal R}}$ is a symmetry of the system. It can be easily verified that $({\hat {\cal G}}{\hat {\cal R}})^4=1$, and therefore the eigenstates of~(\ref{eq:Hamiltonian_Majorana_square}) can be one, two or four-fold degenerate. 

To progress further, we introduce the fermion operator ${\hat f}_{\ell, m}$, such that ${\hat c}_{\ell, m} = i^{\ell+m} {\hat f}_{\ell, m}^\dagger +  (-i)^{\ell+m} {\hat f}_{\ell, m}$ and ${\hat d}_{\ell, m} = i^{\ell+m+1} {\hat f}_{\ell, m}^\dagger +  (-i)^{\ell+m+1} {\hat f}_{\ell, m}$. Plugging this expressions into Eq.~(\ref{eq:Hamiltonian_Majorana_square}) we find
\be \label{eq:Hamiltonian_Majorana_square_2}
&& {\hat {\cal H}_{\rm M}}= \sum_{\ell, m} \big\{
u_{\ell,m}^x \big[(-)^{\ell+m} t_m {\hat f}_{\ell,m}^\dagger {\hat f}_{\ell,m+1} - \Delta_m {\hat f}_{\ell,m}^\dagger {\hat f}_{\ell,m+1}^\dagger \big]
\nn
&&
-
u_{\ell,m}^y \big[(-)^{\ell+m} t_\ell {\hat f}_{\ell,m}^\dagger {\hat f}_{\ell+1,m} - \Delta_\ell {\hat f}_{\ell,m}^\dagger {\hat f}_{\ell+1,m}^\dagger \big]
\big\} + {\rm h.c.},
%\nn
\ee
where $t_\ell = {\tilde J}_{\ell}  + {\tilde J}_{\ell}'$ and $\Delta_\ell = {\tilde J}_{\ell} - {\tilde J}_{\ell}'$. The Hamiltonian~(\ref{eq:Hamiltonian_Majorana_square_2}) is in form identical to that of spinless electrons paired by p-wave superconductivity. Here, however, the pairing can have a vortex structure commensurate to the lattice (see more below). From now on, we choose the gauge-potential configuration $u_{\ell,m}^x = (-1)^{\ell+m}$ and $u_{\ell,m}^y = +1$, that corresponds to having a $\pi$-flux in each plaquette. 

\begin{figure}[t]
\begin{center}
\begin{tabular}{p{0.6\columnwidth}p{0.25\columnwidth}}
\parbox[c]{\hsize}{
\begin{overpic}[width=0.6\columnwidth]{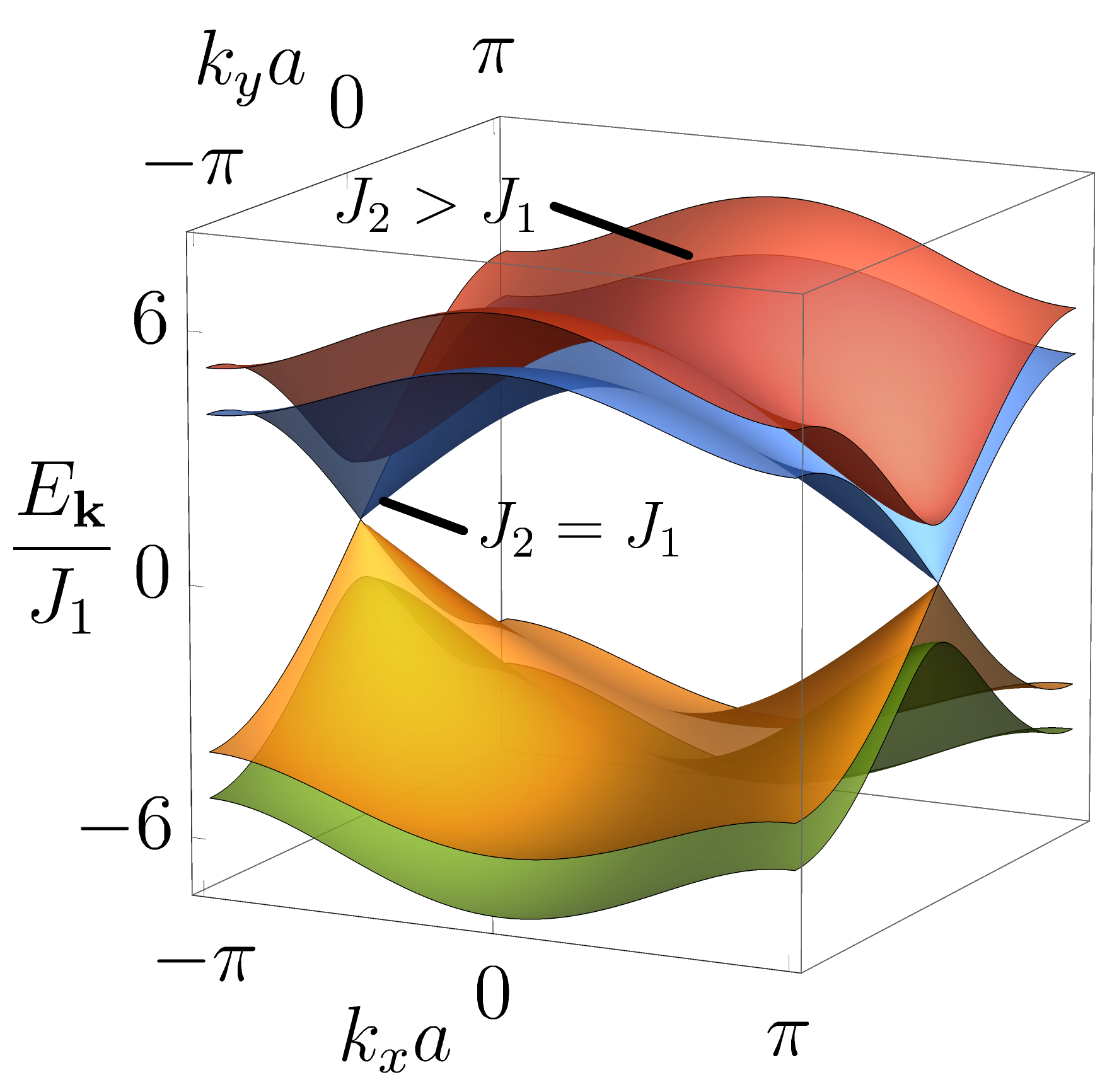}
\put(0,-10){(a)}
\end{overpic}
}
&
\begin{tabular}{c}
\begin{overpic}[width=0.25\columnwidth]{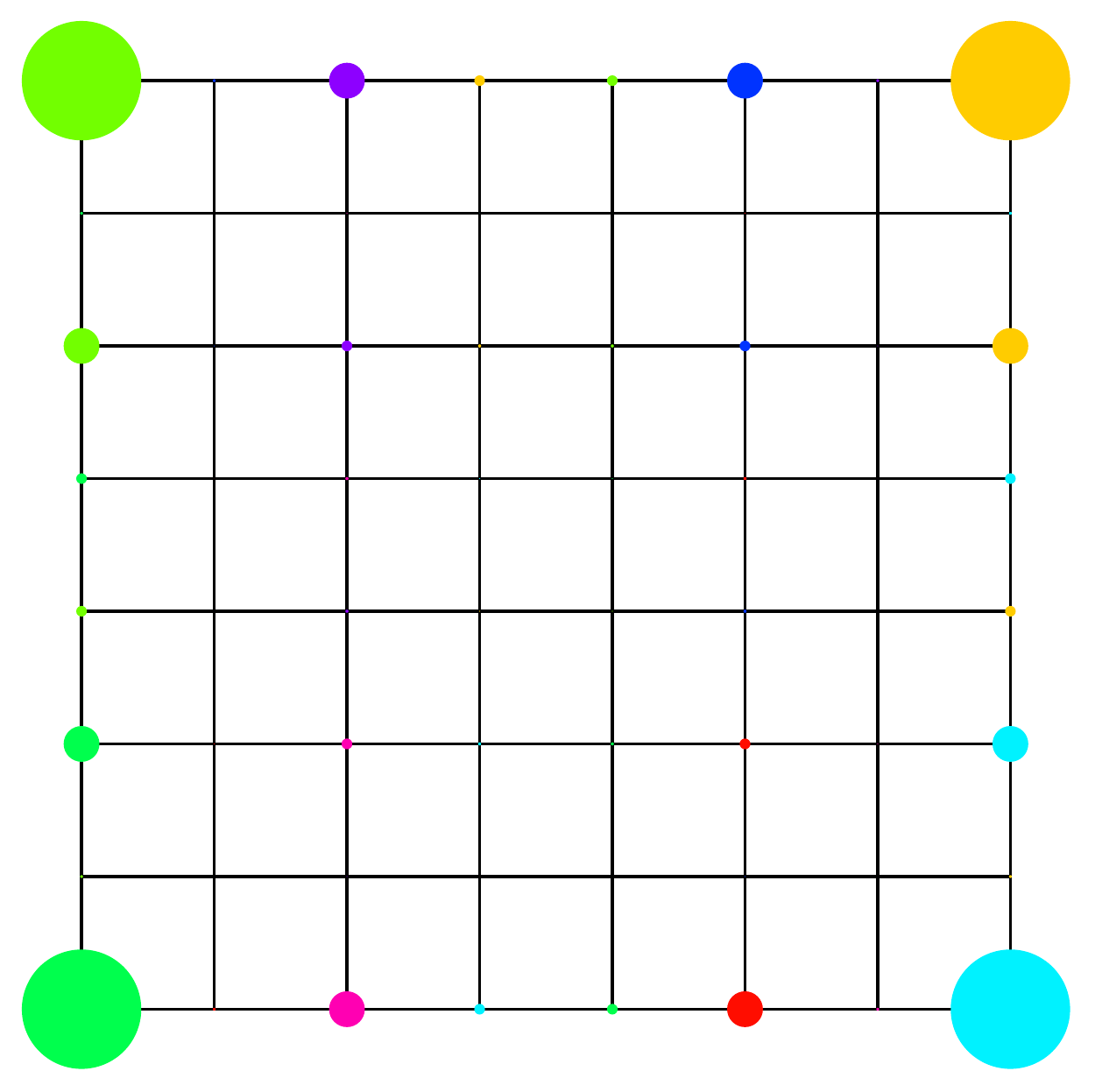}
\put(0,-10){(b)}
\end{overpic}
\vspace{0.3cm}\\
\begin{overpic}[width=0.25\columnwidth]{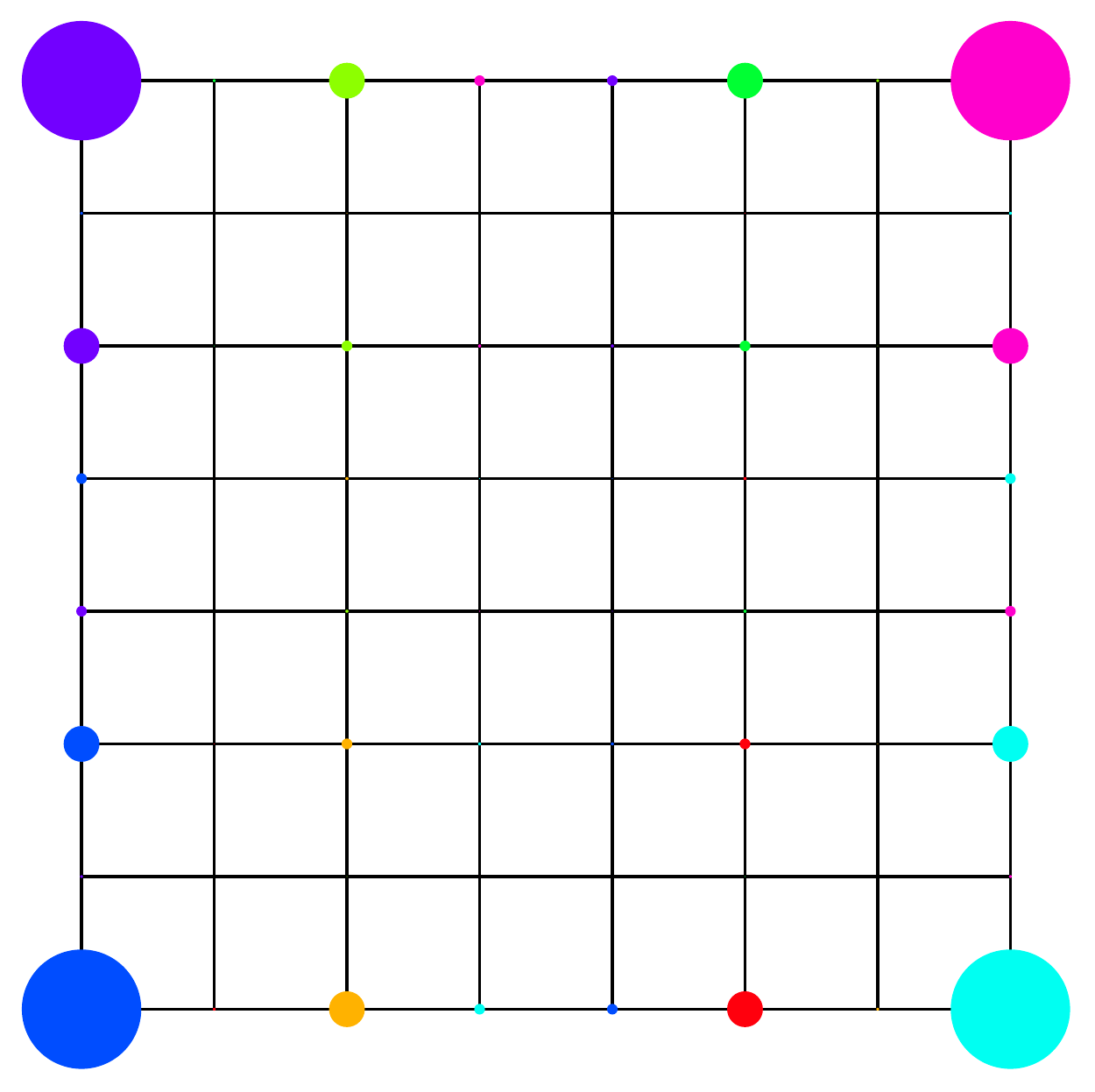}
\put(0,-10){(c)}
\end{overpic}
\vspace{0.4cm}\\
\begin{overpic}[width=0.25\columnwidth]{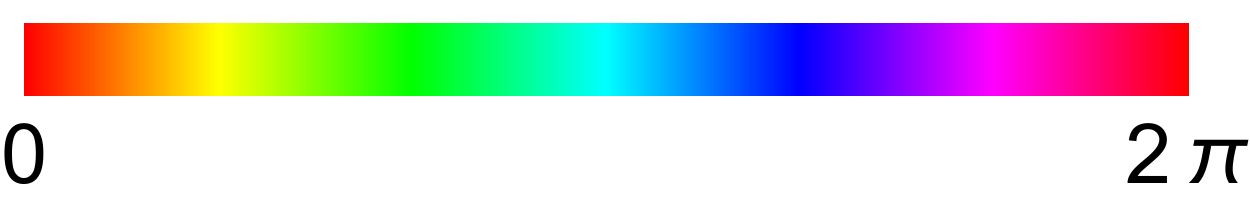}
%\put(0,-10){(d)}
\end{overpic}
\end{tabular}
\end{tabular}
\end{center}
\caption{
Panel (a) The eigenvalues of Hamiltonian~(\ref{eq:square_H_0}) for both $J_2 = J_1$ (inner bands) and $J_2>J_1$ (outer bands).
Panels (b) and (c) The two corner states of Majorana particles obtained by numerically diagonalizing the $c$-part of Hamiltonian~(\ref{eq:Hamiltonian_Majorana_square}) for $J_1=0.3$ and $J_2=1$. 
Due to the structure of Eq.~(\ref{eq:Hamiltonian_Majorana_square}), $d$-particles would exhibit the same states for $J_2'>J_1'$.
}
\label{fig:two}
\end{figure}

Up until now, we have left the couplings $J_1, J_2, J_1'$ and $J_2'$ unspecified. Now we study two interesting situations. In the first case we set $J_1'=J_1$ and $J_2'=J_2$. In turn, $t_\ell = (J_1 + J_2) + (-)^\ell (J_1 - J_2)$ ($t_m$ is analogously defined), while $\Delta_\ell = \Delta_m = 0$. The resulting Hamiltonian is in form identical to that of a fermionic SOTI~\cite{Benalcazar_Science_2017}, and therefore exhibits a phase transition at $J_1 = J_2$. For a system with periodic boundary conditions we can rewrite ${\hat {\cal H}}_{\rm M} = \sum_{{\bm k},\alpha,\beta} {\hat F}_{{\bm k},\alpha}^\dagger H^{(0)}_{{\bm k},\alpha\beta} {\hat F}_{{\bm k},\beta}$, where ${\hat F}_{{\bm k}} = {}^t ({\hat f}_{{\bm k},1}, {\hat f}_{{\bm k},2}, {\hat f}_{{\bm k},3}, {\hat f}_{{\bm k},4})$ [the unit cell, with the four sites $1=1,\ldots, 4$ is shown in Fig.~\ref{fig:one}(a)]. Here, ${}^t$ denotes transposition, while ($a$ is the side of the unit cell)
\be \label{eq:square_H_0}
H^{(0)}_{\bm k} &=& \big[t_0+t_1\cos(k_x a)\big] \openone\otimes  \sigma^x + \big\{ t_1 \sin(k_y a)  \sigma^x
\nn
&+&
\big[t_0-t_1\cos(k_y a)\big] \sigma^y  + t_1 \sin(k_x a)  \sigma^z \big\} \otimes \sigma^y.
%\nn
\ee
Here, $\openone$ is the $2\times 2$ identity matrix and $\sigma^a$ ($a=x,y,z$) are Pauli matrices. Fig.~\ref{fig:two}(a) shows the eigenvalues of Eq.~(\ref{eq:square_H_0}) for two values of $J_1$ and $J_2$. When $J_1 = J_2$, the two bands are doubly-degenerate and touch linearly at the points $(\pm \pi/a,0)$. A gap opens when $J_2\neq J_1$.
When such system is made finite, it is seen to transition from trivial insulator to SOTI~\cite{Benalcazar_Science_2017} at $J_1=J_2$. In Fig.~\ref{fig:two}(b)-(c), we show the lowest eigenstates of Eq.~(\ref{eq:Hamiltonian_Majorana_square_2}), defined on a square lattice with four unit cells along each edge, for $J_2>J_1$. These are zero-energy corner states, each of which is doubly degenerate and exhibits a periodic modulation along the edges reminiscent of SSH edge states. In fact the weights of corner states alternate between finite and zero going along each of the two neighboring edges. 
Edge states also appear at finite energy. Unlike corner states, these are four-fold degenerate. 

More intriguing is the configuration in which $J_1'=J_2$ and $J_2'=J_1$, which imply $t_\ell = t_m = J_1 + J_2$ and $\Delta_\ell = (-)^\ell (J_1 - J_2)$. In this case, both the hopping part and the p-wave pairing in Eq.~(\ref{eq:Hamiltonian_Majorana_square_2}) are characterized by a $\pi$-flux structure. The spin system is therefore mapped into spinless fermions under the simultaneous effect of $\mathbb{Z}_2$ fluxes and p-wave pairing exhibiting a $\pi$-flux structure ({\it i.e.} an Abrikosov lattice of ``half vortices'' commensurate to the square lattice). 
For a system with periodic boundary conditions, the Hamiltonian is
\be \label{eq:Hamiltonian_square_Delta}
{\hat {\cal H}}_{\rm M} = \sum_{\bm k} 
\left(
{\hat F}_{{\bm k}}^\dagger, {\hat F}_{-{\bm k}}
\right)
\left(
\!
\begin{array}{cc}
H^{(0)}_{\bm k} & \Delta_{\bm k}
\vspace{0.1cm}\\
\Delta^\dagger_{\bm k} & -{}^t H^{(0)}_{-{\bm k}}
\end{array}
\!
\right)
\left(
\!\!
\begin{array}{c}
{\hat F}_{{\bm k}} 
\vspace{0.1cm}\\
{\hat F}_{-{\bm k}}^\dagger
\end{array}
\!\!
\right),
%\nn
\ee
where
\be
\Delta_{\bm k} &=&
-i\big[\Delta_0+\Delta_1\cos(k_x a)\big] \openone\otimes  \sigma^y + i \big\{ \Delta_1 \sin(k_y a)  \sigma^x 
\nn
&+&
\big[\Delta_0-\Delta_1\cos(k_y a)\big] \sigma^y  + \Delta_1 \sin(k_x a)  \sigma^z \big\} \otimes \sigma^x.
%\nn
\ee
Since all hopping amplitudes are equal, if the pairing is artificially made to vanish the band structure of Eq.~(\ref{eq:Hamiltonian_square_Delta}) exhibits no gap, exactly as in Fig.~\ref{fig:two}(a). The pairing opens a gap and its vortex structure enables the localization of Majorana particles at the corners of finite systems [analogously to Figs.~\ref{fig:two}(b)-(c)]. Majorana particles are present for all $J_1$ and $J_2$, except when $J_1 = J_2$, for which the gap that stabilizes corner states closes. 

All these findings are made clearer by observing that Eq.~(\ref{eq:Hamiltonian_Majorana_square}) describes two copies of the same ``Majorana SOTI''~\cite{Benalcazar_Science_2017}, one for the ${\hat c}$- and one for ${\hat d}-$particles. When $J_5=0$, the two are independent and exhibit the SOTI transition at $J_1 = J_2$ and $J_1' = J_2'$, respectively. 
When $J_1'=J_1$ and $J_2'=J_2$, both ${\hat c}$- and ${\hat d}$-Majorana particles localize at the corners for $J_2>J_1$ (when the number of plaquettes per side is odd). Therefore, as shown above, corner states have fermionic statistics. 

On the contrary, when $J_1'=J_2$ and $J_2'=J_1$, ${\hat c}$- and ${\hat d}$-Majorana particles localize at the corners on opposite sides of the transition point $J_1 = J_2$. Therefore, for $J_2>J_1$ ($J_2<J_1$) only ${\hat c}$- (${\hat d}$-)Majorana particles localize at the corners. Hence, the system is a Majorana SOTI for all values of $J_1 \neq J_2$. Note that this renders $J_5$ largely ineffective~\footnote{See also the Supplemental Online Material}. Such coupling pairs Majorana particles of different kinds at the same site. However, for any value of $J_1 \neq J_2$ only one type of particle appears at the corners.

\begin{figure}[t]
\begin{center}
\begin{tabular}{ccc}
\begin{overpic}[width=0.31\columnwidth]{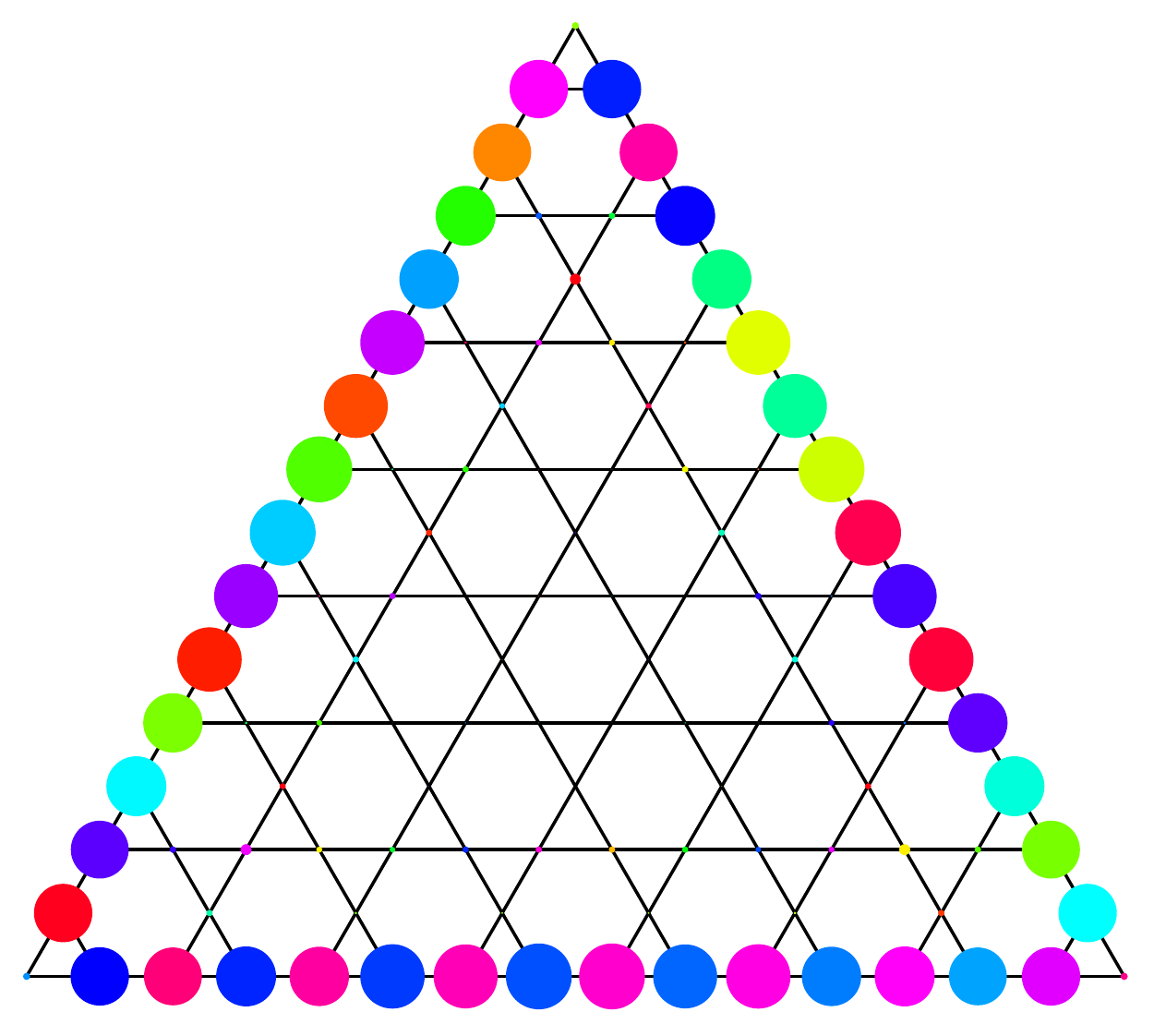}
\put(0,-10){(a)}
\end{overpic}
&
\begin{overpic}[width=0.31\columnwidth]{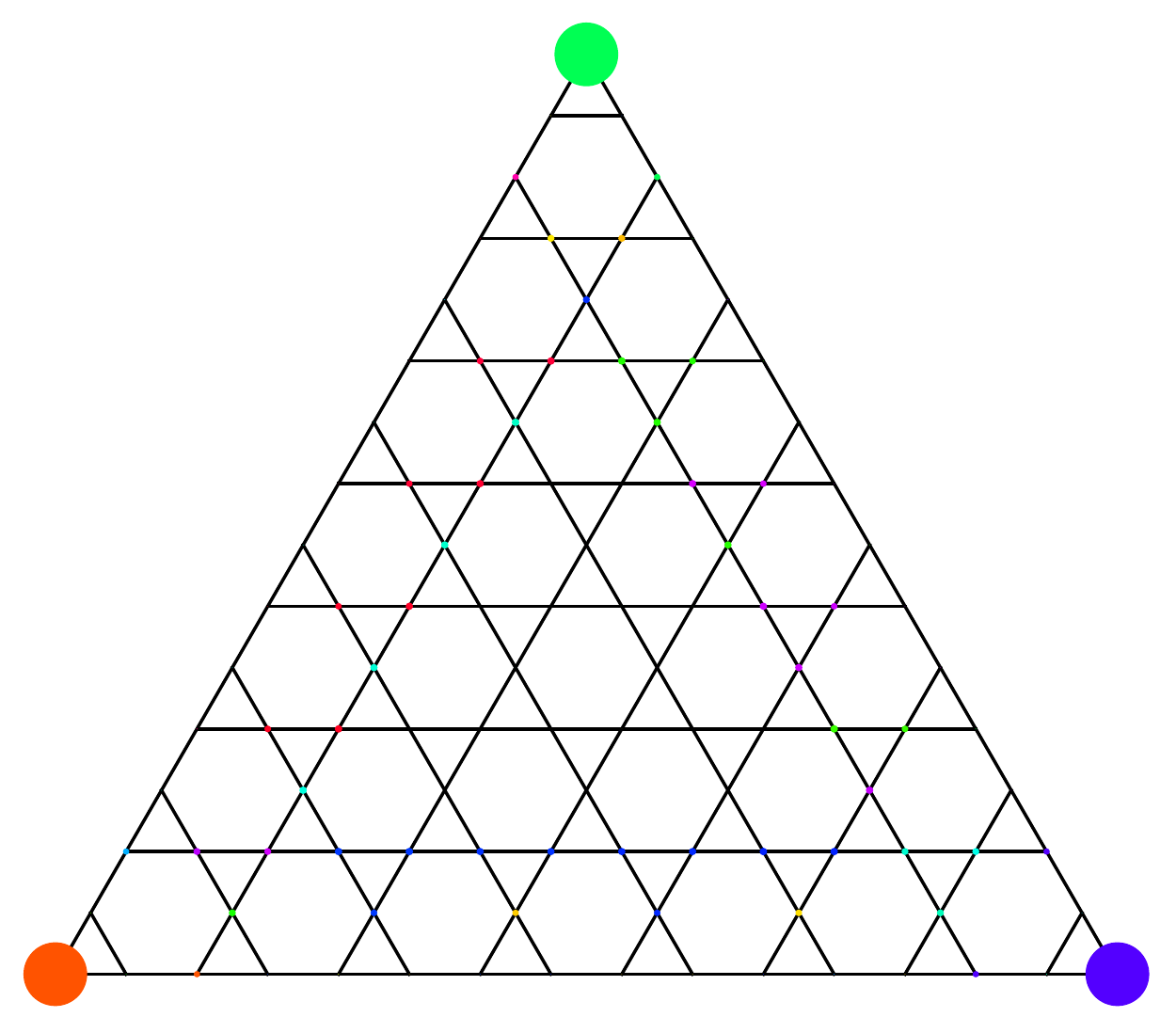}
\put(0,-10){(b)}
\end{overpic}
\vspace{0.5cm}
&
\begin{overpic}[width=0.31\columnwidth]{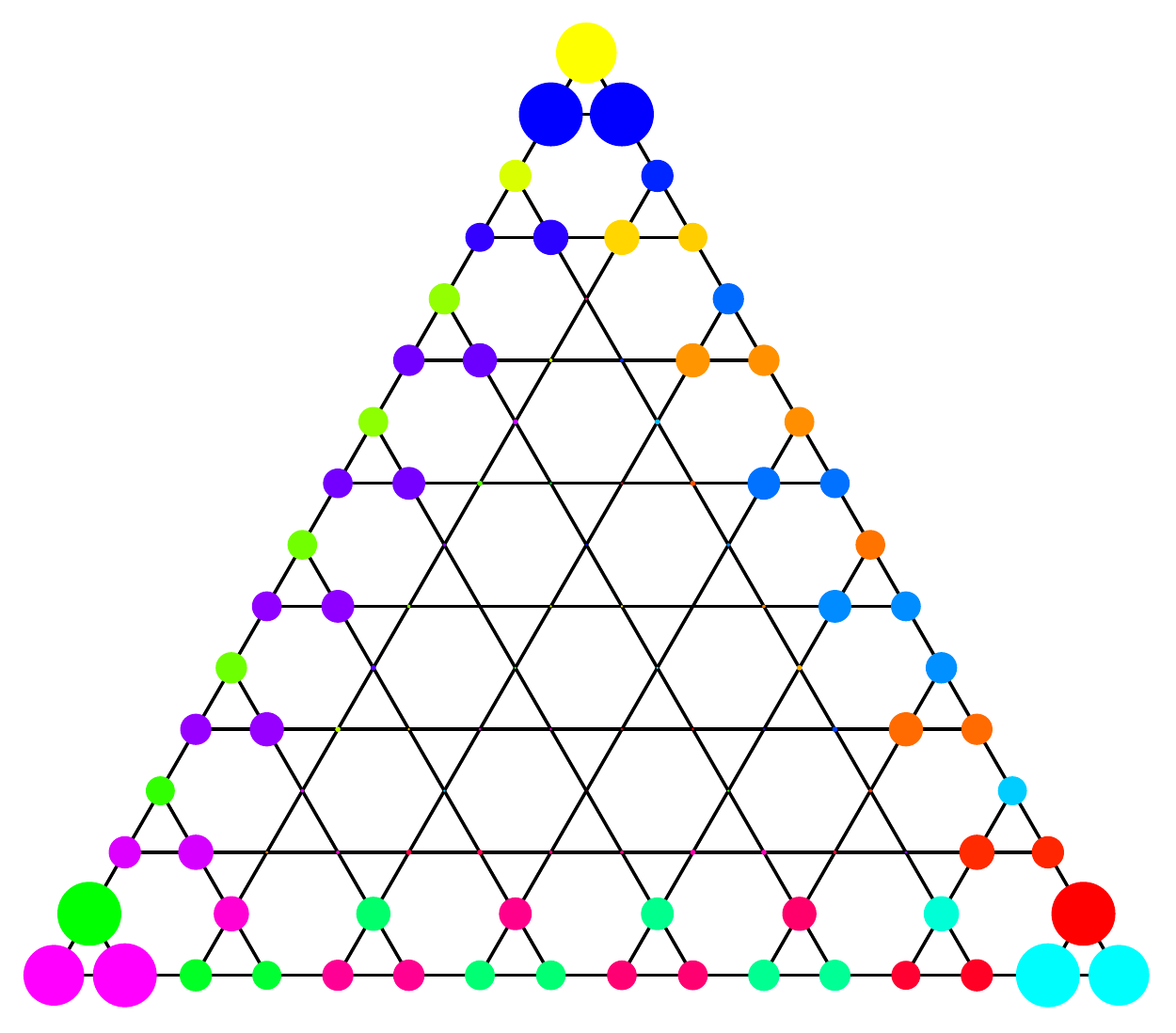}
\put(0,-10){(c)}
\end{overpic}
\end{tabular}
\end{center}
\caption{
Edge and corner states for flux pattern Fig~\ref{fig:one}(e). The color scheme is the same as in Fig.~\ref{fig:two}(b)-(c).
Panel (a) A typical type-one finite energy edge state, obtained with $J_1=0.1$ and $J_2=1$. 
Panel (b) one of the zero-energy corner states, obtained with the same parameters as Panel (a).
Panel (c) Type-two edge state with $J_1=1$ and $J_2=0.1$.
}
\label{fig:three}
\end{figure}

{\it Kagome lattice}---The ground state of the Kagome lattice is not analytically known. For this reason, we explore all four configurations shown in Fig.~\ref{fig:one}(c)-(f). It was indeed shown in Ref.~\cite{Chua_prb_2011} that these have very similar energies, and which of them is the ground state can depend on the size of the lattice. For the sake of presentation, we consider the triangular lattice shown in Fig.~\ref{fig:three}. We diagonalize only the ${\hat c}$-part of Eq.~(\ref{eq:Hamiltonian_Majorana}). In fact, when $J_5=0$ the ${\hat c}$- and ${\hat d}$-particles are completely independent and the behavior of the latter can be deduced by that of the former by replacing $J_1 \to J_1'$ and $J_2 \to J_2'$.

In all four cases we find edge states as in Fig.~\ref{fig:three}(a) coexisting with the zero-energy corner states of Fig.~\ref{fig:three}(b) whenever $J_1<J_2$. We name such states ``type one'', to distinguish them from those of Fig.~\ref{fig:three}(c) (see below). A typical type-one edge state is localized along a series of lines parallel to the edge, and, similar to those of the SSH model, its weights alternate between zero and nonzero values as one moves farther into the bulk. Similarly, the weight of type-one corner states alternates between zero and nonzero values along the two neighboring edges. 

Contrary to all other configurations, that of Fig.~\ref{fig:one}(e) exhibits a second type of edge states, which we name ``type two''. These edge states are considerably more complicated than the type-one discussed above. Firstly, type-two edge states occur when one of $J_1,J_2$ is significantly larger than the other (at $J_1 = J_2$ the system is gapless). Secondly, they are localized along a line of triangles characterized by strong (either $J_1$ or $J_2$) bonds.

{\it Conclusion}---In this manuscript we have studied two spin-$3/2$ models, defined on Kagome~\cite{Chua_prb_2011} and square~\cite{Yao_prl_2009} lattices, that support QSL phases at low temperature. As in Kitaev's work~\cite{Kitaev_AnnPhys_2006}, spins are fractionalized by means of Majorana particles, in this case six of them. Four of them give rise to a $\mathbb{Z}_2$ gauge potential on top of which the remaining two propagate~\cite{Chua_prb_2011,Yao_prl_2009}. In the case of the square lattice, the ground-state $\mathbb{Z}_2$ flux configuration is exactly known thanks to a theorem by Lieb~\cite{Lieb_prl_1994}, and therefore this problem can be solved analytically. On the contrary, since the Kagome lattice is not bipartite, its ground-state flux configuration is unknown. To obviate this problem, we have studied the four possible gauge-inequivalent flux configurations commensurate to the unit cell. 

We find that both models can support topologically-protected corner states when magnetic couplings are made unequal. Such states are fermions when both free particles resulting from the spin fractionalization are able to localize at the corners. On the contrary, when only one of the two species can localize, the corner states are Majorana particles. Notably, in the latter case such quasiparticles are protected against perturbation that locally mix the two kinds of free Majorana particles. We observe that the ground-state energy is lowered by making the magnetic couplings unequal. Therefore, we speculate that real systems could undergo lattice distortions in order to lower their magnetic energy and simultaneously localize fermionic or Majorana states at their corners.

In the Kagome lattice, corner states coexist with edge states, a distinctive trait of the chiral QSL realized in such system~\cite{Chua_prb_2011}. We find that all four flux configurations exhibit the same type of edge and corner states, but one of them supports also a second set of edge states which have no analog in the other three. 

We conclude by noting that corner Majorana states have been predicted to occur in a similar spin-$3/2$ model defined on a Shastry-Sutherland lattice~\cite{Dwivedi_prb_2018}. The latter is akin to our Kagome model: it exhibits both chiral and gapped QSL phases but the ground state is not exactly known since Lieb's theorem does not apply. On the contrary, the ground state flux configuration of the square-lattice model we study is exactly known. We believe that this fact lends credibility to the possibility of localizing fermionic or Majorana particles at corners of QSLs. Furthermore, we hope that the relative simplicity of the lattices studied here and their relative abundance in nature will stimulate the search of material realizations of second-order QSL supporting topologically-protected corner states.

{\it Acknowledgments}---A.P. acknowledges support from the European Commission under the EU Horizon 2020 MSCA-RISE-2019 programme (project 873028 HYDROTRONICS).

\appendix
\section{Protection of edge states against $J_5$}

\begin{figure}[t]
\begin{center}
\begin{tabular}{cc}
\begin{overpic}[width=0.49\columnwidth]{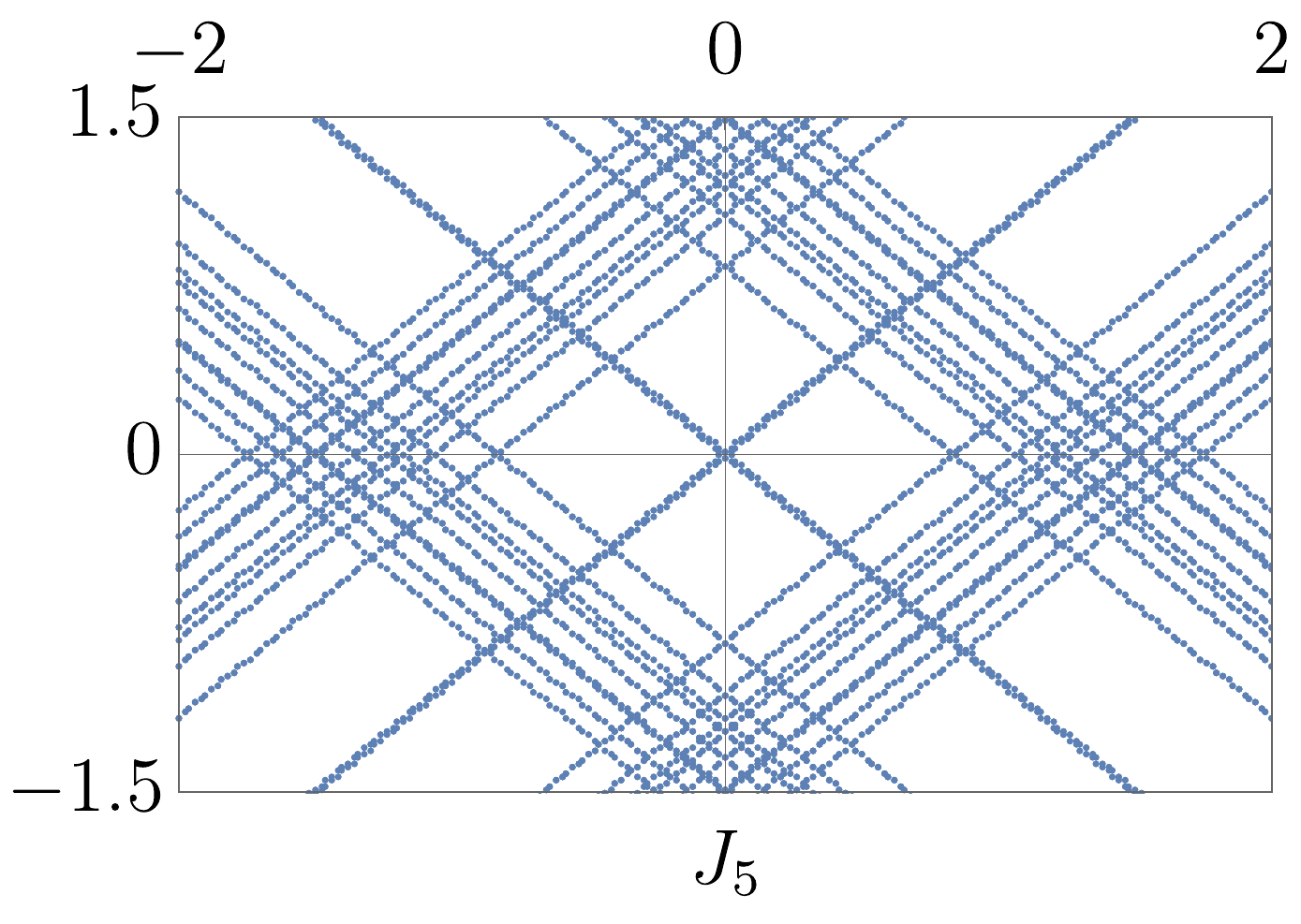}
\put(0,-10){(a)}
\end{overpic}
&
\begin{overpic}[width=0.49\columnwidth]{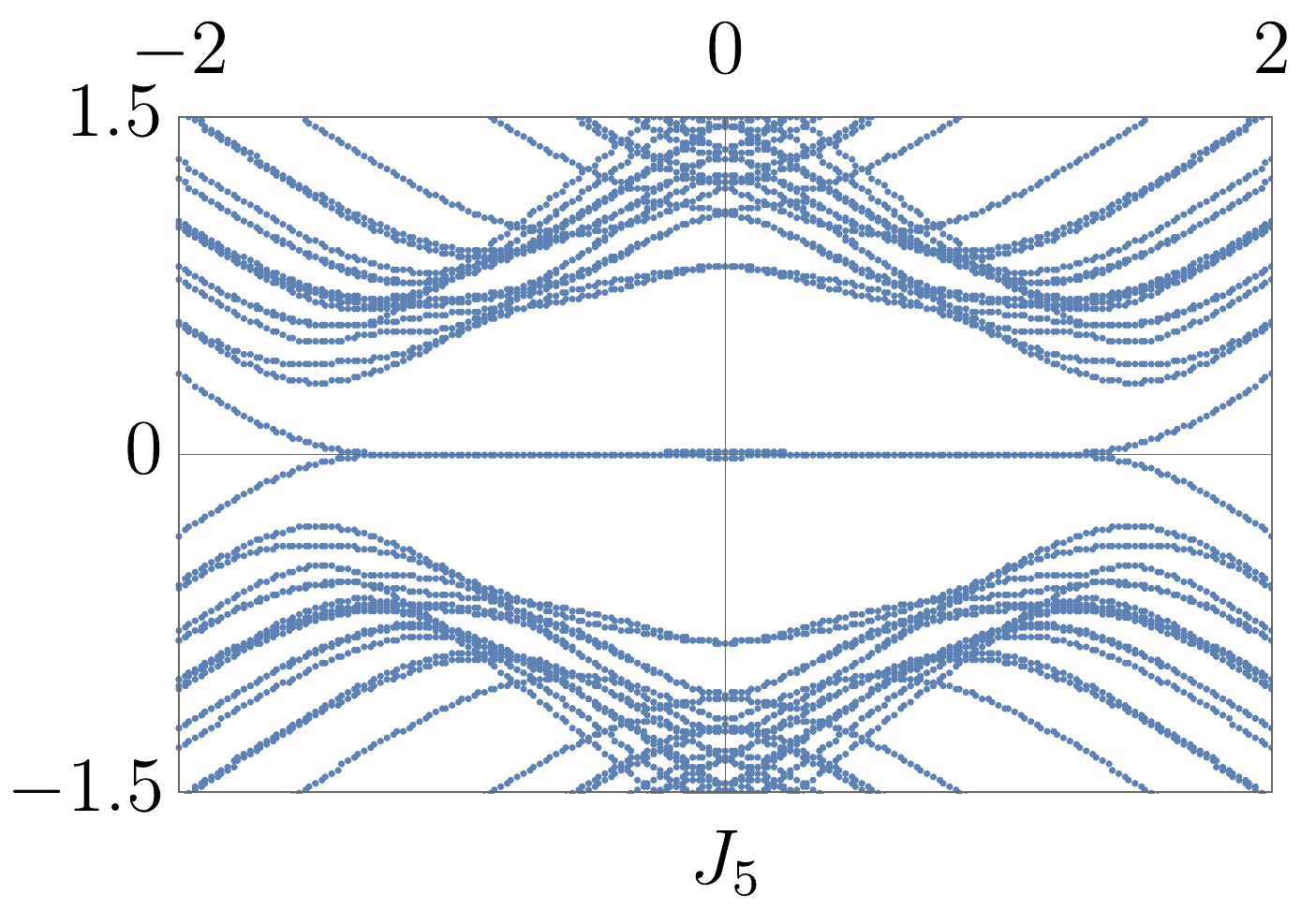}
\put(0,-10){(b)}
\end{overpic}
\end{tabular}
\end{center}
\caption{
Energy spectrum of the square model as a function of $J_5$. In both figures, $J_1=0.3$ and $J_2=1$. 
Panel (a) Spectrum with $J_1'=J_1$ and $J_2'=J_2$. 
Panel (b) Spectrum with $J_1'=J_2$ and $J_2'=J_1$. 
It is evident that Majorana corner states [Panel (b)] are protected against the inter-species coupling $J_5$ up to $|J_5| \sim 1.3$.
}
\label{fig:appendix}
\end{figure}

In this Supplemental Material we show that Majorana corner states are robust against $J_5$ by computing the energy spectrum of the square model for both particle species as a function of $J_5$.

In the case where $J_1'=J_1=0.3$ and $J_2'=J_2=1$ (fermionic corner states), we find that the energies of corner states split away from zero linearly as $J_5$ increases in strength, as shown in Fig.~\ref{fig:appendix}(a).

On the contrary, when $J_1'=J_2=1$ and $J_2'=J_1=0.3$ (Majorana corner states), the corner states remain at zero energy for $|J_5| \lesssim 1.3$. We conclude that on-site mixing the of the two species is forbidden for small-to-moderate inter-species couplings as expected.

\bibliographystyle{apsrev4-1}
\bibliography{biblio}

\end{document}